\documentclass{article}%
\usepackage{graphicx}
\usepackage{amsmath}
\usepackage{amsfonts}
\usepackage{amssymb}
\usepackage{latexsym}
\usepackage{float}
\providecommand{\U}[1]{\protect\rule{.1in}{.1in}}

\newtheorem{theorem}{Theorem}

\newtheorem{definition}[theorem]{Definition}

\newtheorem{remark}[theorem]{Remark}

\newcommand{\mb}[1]{\mathbf{#1}}

\newcommand{\E}{\mathbb{E}}

\begin{document}

\begin{center}
{\Large Symmetric Gini Covariance and  Correlation}

\bigskip

\centerline{\today}
\bigskip   Yongli Sang, Xin Dang and Hailin Sang

\bigskip Department of Mathematics, University of Mississippi,
University, MS 38677, USA. E-mail address: ysang@go.olemiss.edu, xdang@olemiss.edu, sang@olemiss.edu
\end{center}

\begin{center}
\bigskip\textbf{Abstract}
\bigskip
\end{center}
Standard Gini covariance and Gini correlation play important roles in measuring the dependence of random variables with heavy tails. However, the asymmetry brings a substantial difficulty in interpretation. In this paper, we propose a  symmetric Gini-type covariance and a  symmetric Gini correlation ($\rho_g$) based on the joint rank function. The proposed correlation $\rho_g$ is more robust than the Pearson correlation but less robust than the Kendall's  $\tau$ correlation. We establish the relationship
between $\rho_g$ and the linear correlation $\rho$  for a class of random vectors in the family of elliptical distributions, which allows us to estimate $\rho$
based on estimation of $\rho_g$. The asymptotic normality of the resulting estimators of $\rho$ are studied through
two approaches: one from influence function and the other from U-statistics and the delta method. We compare asymptotic efficiencies of linear correlation estimators based on the symmetric Gini, regular Gini, Pearson and Kendall's  $\tau$ under various distributions.
In addition to reasonably balancing between robustness and efficiency, the proposed measure $\rho_g$ demonstrates  superior finite sample performance, which makes it attractive in applications.\\



\section{Introduction}
Let $X$ and $Y$ be two non-degenerate random variables with marginal  distribution functions $F$ and $G$, respectively, and a joint distribution function $H$. To describe dependence correlation between $X$ and $Y$, the Pearson correlation (denoted as $\rho_p$) is probably the most frequently used measure.  This measure is based on the covariance between two variables, which is optimal  for the linear association between bivariate normal variables. However,  the Pearson correlation performs poorly for variables with heavily-tailed or asymmetric distributions, and may be seriously impacted even by a single outlier (e.g.,  Shevlyakov and Smirnov, 2011).  Under the assumption that $F$ and $G$ are continuous,  the Spearman correlation, a robust alternative, is a multiple (twive) of the covariance between the cumulative functions (or ranks) of two variables;
the Gini correlation is based on the covariance between one variable and the cumulative distribution of the other  (Blitz and Brittain, 1964). Two Gini correlations can be defined as  
\begin{align*}
\gamma(X,Y):=\frac{\mbox{cov}(X, G(Y))}{\mbox{cov}(X, F(X))} \;\;\; \text{and} \;\;
\gamma(Y,X):=\frac{\mbox{cov}(Y, F(X))}{\mbox{cov}(Y, G(Y))} 
\end{align*}
to reflect different roles of $X$ and $Y.$
The representation of Gini correlation $\gamma(X,Y)$  indicates that it has mixed  properties of those of the Pearson and Spearman correlations. It is similar to Pearson in $X$ (the variable taken in its variate values) and similar to Spearman in $Y$ (the variable taken in its ranks). Hence Gini correlations complement the Pearson and Spearman correlations (Schechtman and Yitzhaki, 1987; 1999;  2003).  Two Gini correlations are equal if $X$ and $Y$ are exchangeable up to a linear transformation. However, Gini covariances are not symmetric in $X$ and $Y$ in general.  On one hand, this asymmetrical nature is useful and can be used for testing bivariate exchangeability (Schechtman, Yitzhaki and Artsev, 2007). On the other hand,  such asymmetry violates the axioms of correlation
measurement (Mari and Kotz, 2001).   Although some authors (e.g., Xu {\em et al.}, 2010) dealt with asymmetry by a simple average $(\gamma(X,Y)+ \gamma(Y,X))/2$, it is  difficult to interpret this measure,  especially when $\gamma(X,Y)$ and $\gamma(Y,X)$ have different signs.

The asymmetry of $\gamma(X,Y)$ and  $\gamma(Y,X)$ stems from  the usage of marginal rank function $F(x)$ or $G(y)$. A remedy is to utilize a joint rank function. To do so, let us look at a representation of the Gini mean difference (GMD) under continuity assumption:  
$\Delta(F)= 4\mbox{cov}(X, F(X))=2\mbox{cov}(X, 2F(X)-1)$ (Stuart, 1954; Lerman and Yitzhaki, 1984).  The second equality rewrites GMD as twice of the covariance of $X$ and the {\em centered rank} function $r(X):=2F(X)-1$. If $F$ is continuous, $\E r(X)=0$. Hence 
\begin{align}\label{sgmd}
\Delta(F)=2\mbox{cov}(X, r(X))=2\E(Xr(X)).
\end{align}

The rank function $r(X)$ provides a center-orientated ordering with respect to the distribution $F$. Such a rank concept is of vital importance for high dimensions where the natural linear ordering on the real line no longer exists. A generalization of the centered rank in high dimension is called the spatial rank. Based on this joint rank function, 
we are able to propose a symmetric Gini covariance (denoted as $\mbox{cov}_{g}$) and a corresponding symmetric correlation (denoted as $\rho_g$). That is, $\mbox{cov}_{g}(X,Y)=\mbox{cov}_{g}(Y,X)$ and $\rho_g(X,Y)=\rho_g(Y,X)$.

We study properties of the proposed Gini correlation $\rho_g$. In terms of the influence function,  $\rho_g$ is more robust than the Pearson correlation $\rho_p$. However, $\rho_g$ is not as robust as the Spearman correlation and Kendall's $\tau$ correlation.  Kendall's $\tau$ is another commonly used nonparametric measure of association. The Kendall correlation measure is more robust and more efficient than the Spearman correlation (Croux and Dehon, 2010). For this reason, in this paper we do not consider Spearman correlation for comparison.  

As Kendall's $\tau$ has a relationship with  the linear correlation $\rho$ under elliptical distributions (Kendall and Gibbons, 1990; Lindskog, Mcneil and Schmock, 2003), we also set up a function between $\rho_g$ and $\rho$ under elliptical distributions. This provides us an alternative method to estimate $\rho$ based on estimation of $\rho_g$. The asymptotic normality of the estimator based on the symmetric Gini correlation is established. Its asymptotic efficiency and finite sample performance are compared with those of Pearson, Kendall's $\tau$ and the regular Gini correlation coefficients under various  elliptical distributions. 

As any quantity based on spatial ranks, $\rho_g$ is only invariant under translation and homogeneous change. In order to gain the invariance property under heterogeneous changes, we provide an affine invariant version.  

The paper is organized as follows. In Section 2, we introduce a  symmetric Gini covariance and the corresponding correlation.   Section 3 presents the influence function. Section 4 gives an estimator of the symmetric Gini correlation and establishes its asymptotic properties.  In Section 5, we present the affine invariant version of the symmetric Gini correlation and explore finite sample efficiency of the corresponding estimator. We present a real data application of the proposed correlation in Section 6.  Section 7 concludes the paper with a brief summary. All proofs are reserved to the Appendix.
\section{Symmetric Gini covariance and correlation}
The main focus of this section is  to present the proposed symmetric  Gini covariance and correlation, and to study the corresponding properties. 


\subsection{Spatial rank} 

Given a random vector $\mb Z$ in $\mathbb{R}^d$ with distribution $H$, the {\em spatial rank} of $\mb z$ with respect to the distribution $H$ is defined as
\begin{equation*} \label{eqn:sr}
\mb r(\mb z,H)
:=\E \mb s(\mb z-\mb Z)
=\E \frac{\mb z-\mb Z}{||\mb z-\mb Z||},
\end{equation*}
where $\mb s(\cdot)$ is the {\em spatial sign} function defined as $\mb s(\mb z)=\mb z/\|\mb z\|$ with $\mb s(\mb{0})=\mb{0}$.  The solution of $ \mb r(\mb z,H)=\mb 0$ is called the {\em spatial median} of $H$, which minimizes $\E_H\|\mb z-\mb Z\|$.  Obviously, $\E \mb r(\mb Z,H)=\mb 0$ if $H$ is continuous. For more comprehensive account on the spatial rank, see  Oja (2010). 

In particular,  for $d=2$ with $\mb Z=(X,Y)^T$, the bivariate spatial rank function of $\mb z=(x,y)^T$ is 
\begin{equation*}
\mb r(\mb z,H)
=\E \displaystyle\frac{ (x-X,y-Y)^T}{\|\mb z-\mb Z\|}:=(R_1(\mb z), R_2(\mb z))^T,
\end{equation*}
where $R_1(\mb z)=\E (x-X)/\|\mb z-\mb Z\|$ and $R_2(\mb z)=\E(y-Y)/\|\mb z-\mb Z\|$ are two components of the joint rank function $\mb r(\mb z,H)$.
\subsection{Symmetric Gini covariance}
Our new symmetric covariance and correlation are defined based on the bivariate spatial rank function.
Replacing the univariate centered rank in (\ref{sgmd}) with $R_2(\mb z)$, we define
the {\em symmetric  Gini  covariance} as
\begin{align}
 \mbox{cov}_{g}(X, Y):=2 \E X R_{2}(\mb Z).\label{nc}
\end{align}

Note that $\mbox{cov}_{g}(X, Y)=2 \mbox{cov}( X, R_{2}(\mb Z))$ if $H$ is continuous. Dually, $ \mbox{cov}_{g}(Y, X)=2 \E Y R_{1}(\mb Z)$
can also be taken as the definition of the symmetric Gini covariance between $X$ and $Y$. Indeed, 
\begin{align}
& \mbox{cov}_{g}(X, Y)=2\E X R_{2}(\mb Z)=2\E (X_1 \E \big [\frac{ Y_1-Y_2}{||\mb Z_1-\mb Z_2||}\big|\mb Z_1\mb])
=2\E X_1 \frac{ Y_1-  Y_2}{||\mb Z_1-\mb Z_2||}\nonumber \\
&=-2\E X_2 \frac{ Y_1-Y_2}{||\mb Z_1-\mb Z_2||}= \E [ \frac{(X_1-X_2) (Y_1-  Y_2)}{||\mb Z_1-\mb Z_2||}] = \mbox{cov}_{g}(Y, X), \label{gcov}
\end{align}
where $\mb Z_1=(X_1,Y_1)^T$ and $\mb Z_2=(X_2,Y_2)^T$ are independent copies of $\mb Z=(X,Y)^T$ from $H$.
In addition, we define
\begin{align}
&\mbox{cov}_{g}(X, X):=2\E X R_{1}(\mb Z)= \E \frac{(X_1-X_2)^2}{\|\mb Z_1-\mb Z_2\|}; \;\; \label{sgvx}\\
&\mbox{cov}_{g}(Y, Y):=2 \E Y R_{2}(\mb Z)= \E  \frac{(Y_1-Y_2)^2}{\|\mb Z_1-\mb Z_2\|}. \label{sgvy}
\end{align}
We see that not only the Gini covariance between $X$ and $Y$ but also Gini variances of $X$ and of $Y$ are defined jointly through the spatial rank. *** (2015) considered the Gini covariance matrix
$\mb \Sigma_g=2\E\mb Z \mb r^T(\mb Z)$.
The covariances defined above in (\ref{nc}), (\ref{sgvx}) and (\ref{sgvy}) are elements of $\mb \Sigma_g$ for two dimensional random vectors.
Rather than the assumption on a finite second moment in the usual covariance and variance, the Gini counterparts  assume only the first moment, hence being more suitable for heavy-tailed distributions. A related covariance matrix is spatial sign covariance matrix (SSCM), which requires a location parameter to be known but no assumption on moments (Visuri, Koivunen and Oja,  2000). 

Particularly, if $ Z$ is a one dimensional random variable, we have $\mbox{cov}_{g}(Z, Z)=  \E | Z_1-Z_2|$,
which reduces to GMD. In this sense, we may  view the symmetric Gini covariance as a direct generalization of GMD to two variables.


\subsection{Symmetric Gini correlation} \label{sec:sgc}

Using the symmetric Gini covariance defined by (\ref{nc}), we propose a symmetric {\it Gini correlation  coefficient} as follows. 
\begin{definition} \label{def:gcr}
$\mb Z=(X,Y)^T$  is a bivariate random vector from the distribution $H$ with finite first moment and non-degenerate marginal distributions, then the symmetric Gini correlation between $X$ and $Y$ is
\begin{align}
\rho_g(X,Y)&:=\frac{\mbox{cov}_{g}(X, Y)}{\sqrt{ \mbox{cov}_{g}(X, X)}\sqrt{  \mbox{cov}_{g}(Y, Y)}}
=\frac{\E X R_2(\mb Z)}{\sqrt{\E X R_1(\mb Z)}\sqrt{\E Y R_2(\mb Z)}}\label{grhod}.
\end{align}
\end{definition}


\begin{theorem}{}{}\label{corp}
For a bivariate random vector $(X,Y)^T$ from $H$ with finite first moment, $\rho_g$ has the following properties:

\begin{enumerate}
\item $\rho_g(X,Y)=\rho_g(Y,X)$.
\item $-1\leq \rho_g(X,Y) \leq 1$.
\item If $X$, $Y$ are independent, then $\rho_g(X,Y)=0$.
\item If $Y=aX+b$ and $a\ne 0$, then  $\rho_g=\mbox{sgn}(a)$.
\item $\rho_g(aX+b, aY+d)=\rho_g(X,Y)$ for any constants $b$, $d$ and $a\neq 0$.  Measure $\rho_g$ is sensitive to a heterogeneous change, i.e., $\rho_g(aX,cY)\neq \rho_g(X,Y)$ for $a\neq c$. In particular, $\rho_g(X,Y)=-\rho_g(aX,-aY)=-\rho_g(-aX,aY)$.
\end{enumerate}
\end{theorem}

The proof is placed in the Appendix.  Theorem \ref{corp} shows that the symmetric Gini correlation has all properties of Pearson correlation coefficient except for Property 5.  It loses the invariance property under heterogeneous changes because of the Euclidean norm in the spatial rank function. To overcome this drawback, we give the affine invariant version of the $\rho_g$ in Section 5. Comparing with Pearson correlation, as we will see in Section 3, the Gini correlation is more robust in terms of the influence function. 

\subsection{Symmetric Gini correlation for elliptical distributions}

The relationship between Kendall's $\tau$ and the linear correlation coefficient $\rho$, $\tau=2/\pi \arcsin (\rho)$, holds for all elliptical distributions. So $\rho=\sin(\pi \tau/2)$ provides a robust estimation method for $\rho$ by estimating $\tau$ (Lindskog {\em et al.}, 2003). 
This motivates us to explore the relationship between  the symmetric Gini correlation $\rho_g$ and the linear correlation coefficient $\rho$ under  elliptical distributions.

A $d$-dimensional continuous random vector $\mb Z$
has an elliptical distribution if its density function is of the form 
\begin{align}
f(\mb z|\mb\mu, \mb\Sigma)=|\mb\Sigma|^{-1/2}g\{(\mb z-\mb\mu)^{T}\mb\Sigma^{-1}(\mb z-\mb\mu)\}, \label{pdf}
\end{align}
where $\mb\Sigma$ is the scatter matrix, $\mu$ is the location parameter and the nonnegative function $g$ is the density generating function.  An important property for the elliptical distribution is that the nonnegative random variable $R=||\mb\Sigma^{-1/2}(\mb Z-\mb{\mu})||$ is independent of $\mb U=\{\mb\Sigma^{-1/2}(\mb Z-\mb\mu)\}/ R$, which is uniformly distributed on the unit sphere.  When $d=1$, the class of  elliptical distributions coincides with the location-scale class. For $d=2$, let $\mb Z=(X,Y)^T$ and $\Sigma_{ij}$ be the $ (i,j)$ element of $\mb\Sigma$, then the {\em linear correlation coefficient} of $X$ and $Y$ is $\rho=\rho(X,Y):=\frac{\Sigma_{12}}{\sqrt{\Sigma_{11}\Sigma_{22}}}.$ 
If the second moment of $\mb Z$ exists, then the scatter parameter $\mb\Sigma$ is proportional to the covariance matrix. Thus the Pearson correlation $\rho_p$ is well defined and is equal to the parameter $\rho$ in the elliptical distributions. 
If $\Sigma_{11}=\Sigma_{22}=\sigma^2$, we say $X$ and $Y$ are homogeneous, and $\mb\Sigma$ can then be written as $\mb\Sigma=\sigma^2\begin{pmatrix}1 \;\;\;&\rho\\ \rho\;\;\;&1\end{pmatrix}$. In this case, if $\rho=\pm1$, $\mb\Sigma$ is singular and the distribution reduces to an  one-dimensional distribution.

The following theorem states the relationship between $\rho_g$ and $\rho$ under elliptical distributions. 
\begin{theorem}{}{}\label{thm:rel}
If $\mb{Z}=(X,Y)^T$ has an elliptical distribution $H$ with finite first moment and the scatter matrix $\mb\Sigma=\sigma^2\begin{pmatrix}1 \;\;\;&\rho\\ \rho\;\;\;&1\end{pmatrix}$, then we have 
\begin{align}\label{relation}
&\rho_{g}= k(\rho)=\begin{cases}
      \rho & \rho=0,\pm 1, \\
     \displaystyle\frac{1}{\rho}+\frac{\rho-1}{\rho} \frac{\text{EK}(\frac{2\rho}{\rho+1})}{\text{EE}(\frac{2\rho}{\rho+1})}, & otherwise,
   \end{cases}
\end{align}
where
\begin{align*}
\text{EK}(x)=\int_{0}^{\pi/2}\frac{1}{\sqrt{1-x^2 \sin^{2}\theta}} \; d\theta \;\;\mbox{ and }\;\;\text{EE}(x)=\int_{0}^{\pi/2}\sqrt{1-x^2 \sin^{2}\theta} \;  d\theta
\end{align*}
are the complete elliptic integral of the first kind and the second kind, respectively. 
\end{theorem}
\begin{figure*}[thb]
\center
\includegraphics[height=4.0in, width=4.0in]{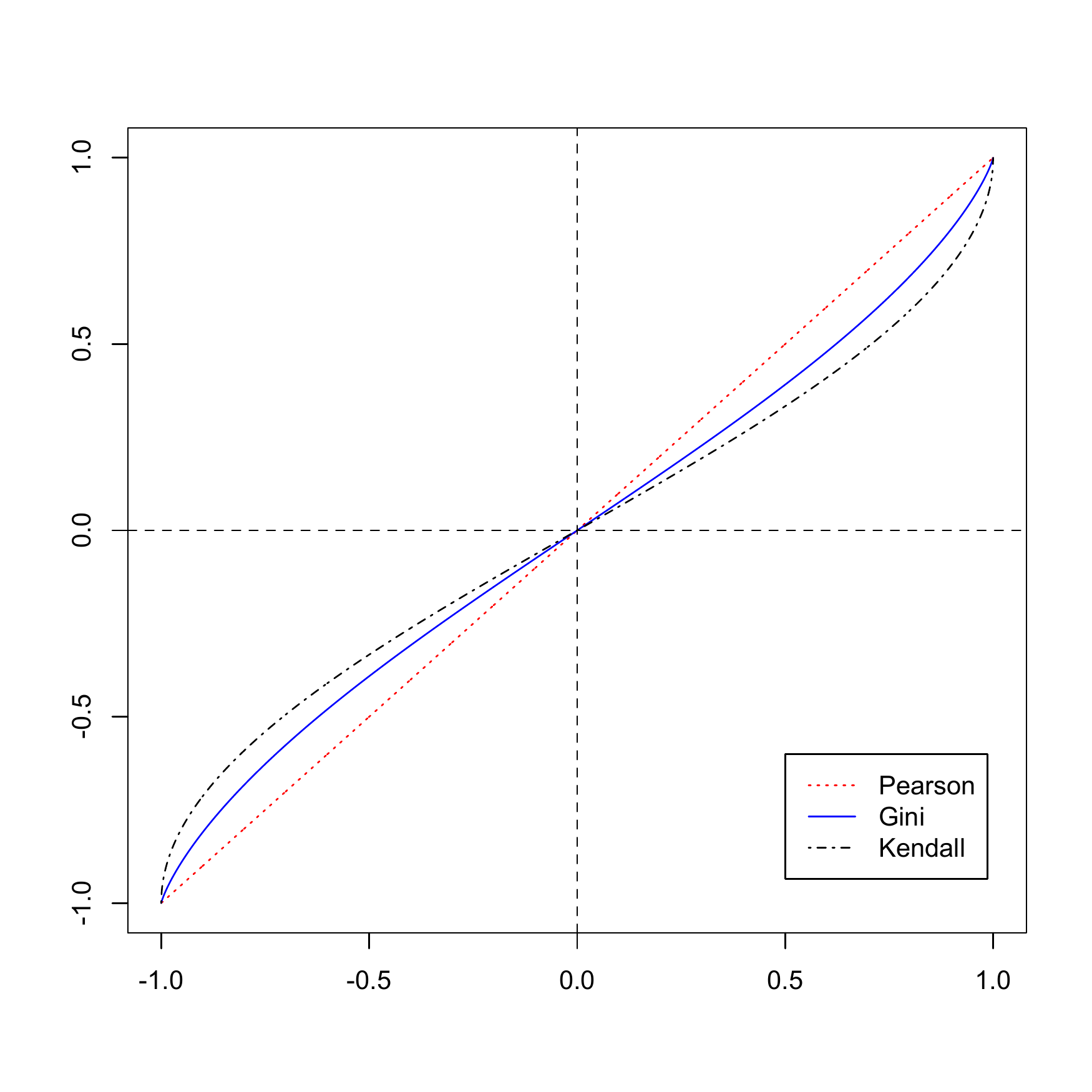}
\vspace{-0.4in}
\caption{Pearson $\rho_p$, Kendall's $\tau$ and symmetric Gini $\rho_g$ correlation coefficients versus $\rho$, the correlation parameter of homogeneous elliptical distributions with finite second moment.}
\label{fig:cor}
\end{figure*} 

The relationship (\ref{relation}) holds only for $\mb\Sigma$ with $\Sigma_{11}=\Sigma_{22}$ because of the loss of invariance property of $\rho_g$ under the heterogeneous changes. 
Note that for any elliptical distribution, the regular Gini correlations are equal  to $\rho$. Schechtman and Yitzhaki (1987) proved that $\gamma(X,Y) =\gamma(Y,X) =\rho$ for bivariate normal distributions, but their proof can be modified for all elliptical distributions.  Based on the spatial sign covariance matrix,  D\"{u}rre, Vogel and Fried (2015) considered a spatial sign correlation coefficient, which equals to $\rho$ for elliptical distributions.

Figure \ref{fig:cor} plots the proposed symmetric Gini correlation $\rho_g$ as  function of $\rho$ under homogeneous elliptical distributions with finite second moment. In comparison, we also plot Pearson $\rho_p$ and Kendall's $\tau$ against $\rho$.  All correlations are increasing in $\rho$.  It is clear that $|\tau|<|\rho_g|<|\rho_p|=|\rho|$. 

With (\ref{relation}), the estimate $\hat{\rho}_g$ of $\rho_g$ can be corrected to ensure Fisher consistency by using the inversion transformation $k^{-1}(\hat \rho_g)$, denoted as $\hat{\rho}^g$.  In the next section, we study the influence function of $\rho_g$, which can be used to evaluate robustness and efficiency of the estimators $\hat{\rho}_g$ in any distribution and that of $\hat{\rho}^g$ under elliptical distributions.

%
%
%
\section{Influence function}\label{influenceF}
The influence function (IF) introduced by Hampel (1974) is now a standard tool in robust statistics
for measuring effects on estimators due to infinitesimal perturbations of sample distribution functions (Hampel {\em et al.}, 1986). For a cdf $H$ on $\mathbb{R}^d$ and a functional $T:H \mapsto T(H) \in \mathbb{R}^m$ with $m\geq 1$, the IF of $T$ at $H$ is defined as
$\mbox{IF}(\mb{z};T,H)
=\displaystyle \lim_{\varepsilon \downarrow 0}
\frac{T((1-\varepsilon)H+\varepsilon\delta_{\mb{z}})-T(H)}{\varepsilon},
\;\;\;\mb{z}\in\mathbb{R}^d,
$
where $\delta_{\mb{z}}$ denotes the point mass distribution at $\mb{z}$. Under regularity conditions on $T$ (Hampel {\em et al.}, 1986; Serfling, 1980), we have $\E_H\{\mbox{IF}(\mb{Z};T,H)\} = \mb 0$ and the von Mises expansion
\begin{equation}\label{ST}
T(H_n)-T(H)= \frac{1}{n}\sum_{i=1}^n\mbox{IF}(\mb{z}_i;T,H)+o_p(n^{-1/2}),
\end{equation}
where $H_n$ denotes the empirical distribution based on sample $\mb{z}_1$,...,$\mb{z}_n$. This representation shows the connection of the IF with robustness of $T$, observation by observation.
Furthermore, (\ref{ST}) yields asymptotic $m$-variate normality of $T(H_n)$,
\begin{equation}\label{ST2}
\sqrt{n}(T(H_n)-T(H)) \stackrel{d}{\rightarrow}
N({\bf 0}, \E_H(\mbox{IF}(\mb{Z};T,H)\mbox{IF}(\mb{Z};T,H)^T) \;\;\text{as $n\to \infty$}.
\end{equation}

To find the influence function of the symmetric Gini correlation defined in (\ref{grhod}),
let $T_1(H)=2\E X R_1(\mb Z)$, $T_2(H)=2\E XR_2(\mb Z)$, $T_3(H)=2\E YR_2(\mb Z)$ and $h(t_1,t_2,t_3)=t_2/{\sqrt{t_1t_3}}$. Then $\rho_g=T(H)=h(T_1,T_2,T_3)$. Denote the influence function of $T_i$ as $L_i(x,y)= \mbox{IF}((x,y)^T; T_i,H)$ for $i=1,2,3$. 
\begin{theorem}{}{}\label{IF}
For any distribution $H$ with finite first moment, the influence function of $\rho_g=T(H)$ is given by 
\begin{align*}
 \mbox{IF}((x,y)^T; \rho_g, H)=&-\frac{\rho_g}{2}\bigg(\frac{L_1(x,y)}{T_1}-\frac{2L_2(x,y)}{T_2}+\frac{L_3(x,y)}{T_3}\bigg) \\
 =& -\frac{\rho_g}{2}\bigg(\frac{1}{T_1}\int\frac{2(x-x_1)^2}{\sqrt{(x-x_1)^2+(y-y_1)^2}}dH(x_1,y_1)\\
 & -\frac{1}{T_2}\int\frac{4(x-x_1)(y-y_1)}{\sqrt{(x-x_1)^2+(y-y_1)^2}}dH(x_1,y_1)\\
&+\frac{1}{T_3}\int\frac{2(y-y_1)^2}{\sqrt{(x-x_1)^2+(y-y_1)^2}}dH(x_1,y_1)\bigg).
\end{align*}
\end{theorem}
Note that each of $L_i(x,y)$ is approximately linear in $x$ or $y$. 
Comparing with the quadratic effects in the Pearson's correlation coefficient (Devlin, Gnanadesikan and Kettering, 1975), 
$$  \mbox{IF}((x,y)^T; \rho_p,H)=\frac{(x-\mu_X)(y-\mu_Y)}{\sigma_X\sigma_Y}-\frac{1}{2}\rho\left[\frac{(x-\mu_X)^2}{\sigma_X^2}+\frac{(y-\mu_Y)^2}{\sigma_Y^2}\right],$$
$\rho_g$ is more robust than the Pearson correlation. However, $\rho_g$ is not as robust as Kendall's $\tau$ correlation since the influence function of $\rho_g$ is unbounded.  Kendall's $\tau$ correlation has a bounded influence function (Croux and Dehon, 2010), which is
$ \mbox{IF}((x,y)^T; \tau, H) = 2\{2P_H[(x-X)(y-Y)>0]-1-\tau\}.$
In this sense, $\rho_g$ is more robust than $\rho_p$ but less robust than  $\tau$. 

\begin{figure*}[h]
\centering
\begin{tabular}{ccc}\vspace{-0.6in}
IF of $\rho_p$ & IF of $\rho_g$ & IF of $\tau$ \\ 
\hspace{-0.2in}\includegraphics[width=1.88in,height=2.7in]{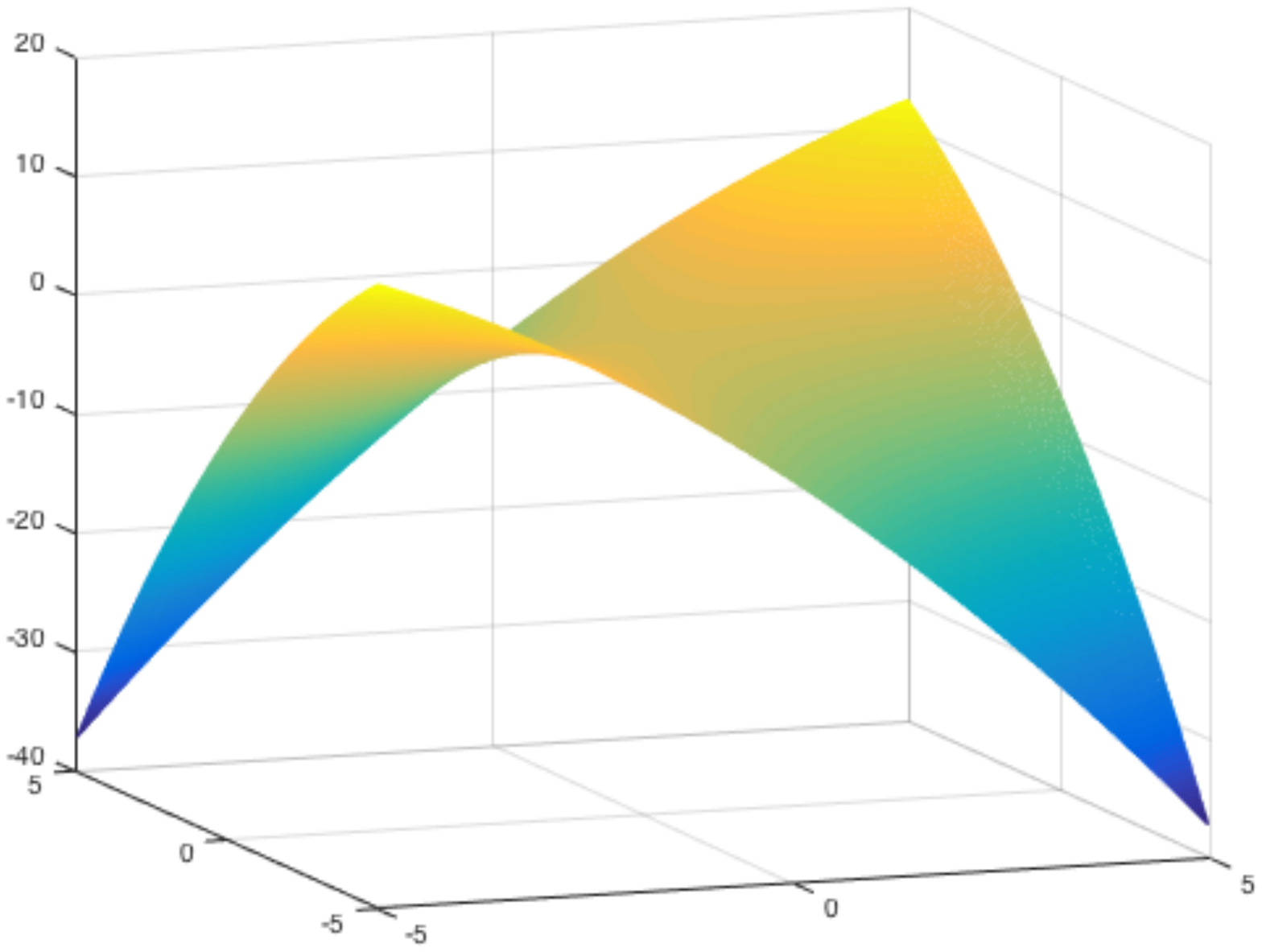} \vspace{-0.25in}&
\hspace{-0.2in}\includegraphics[width=1.88in,height=2.7in]{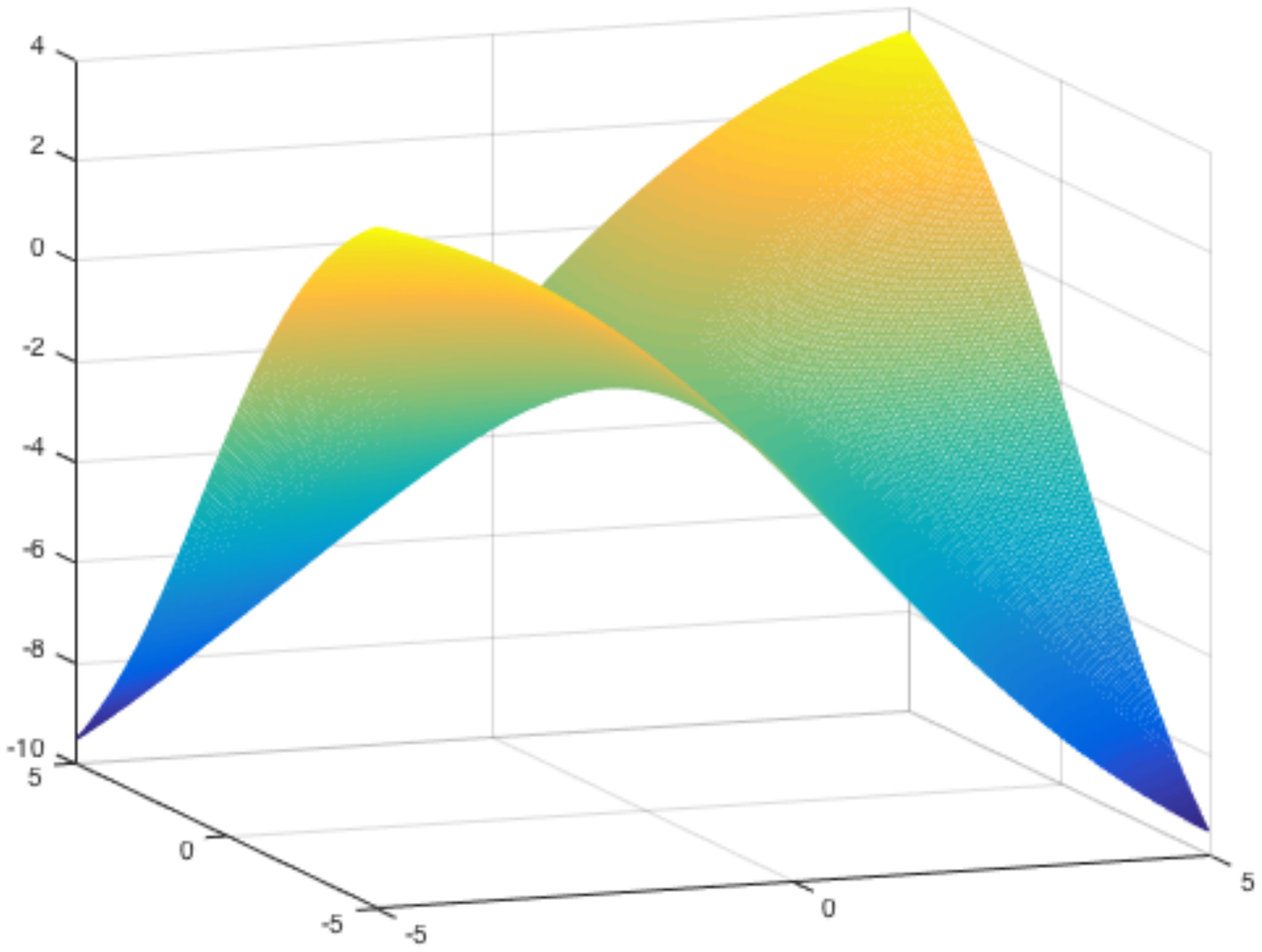} \vspace{-0.25in}&
\hspace{-0.2in}\includegraphics[width=1.88in,height=2.7in]{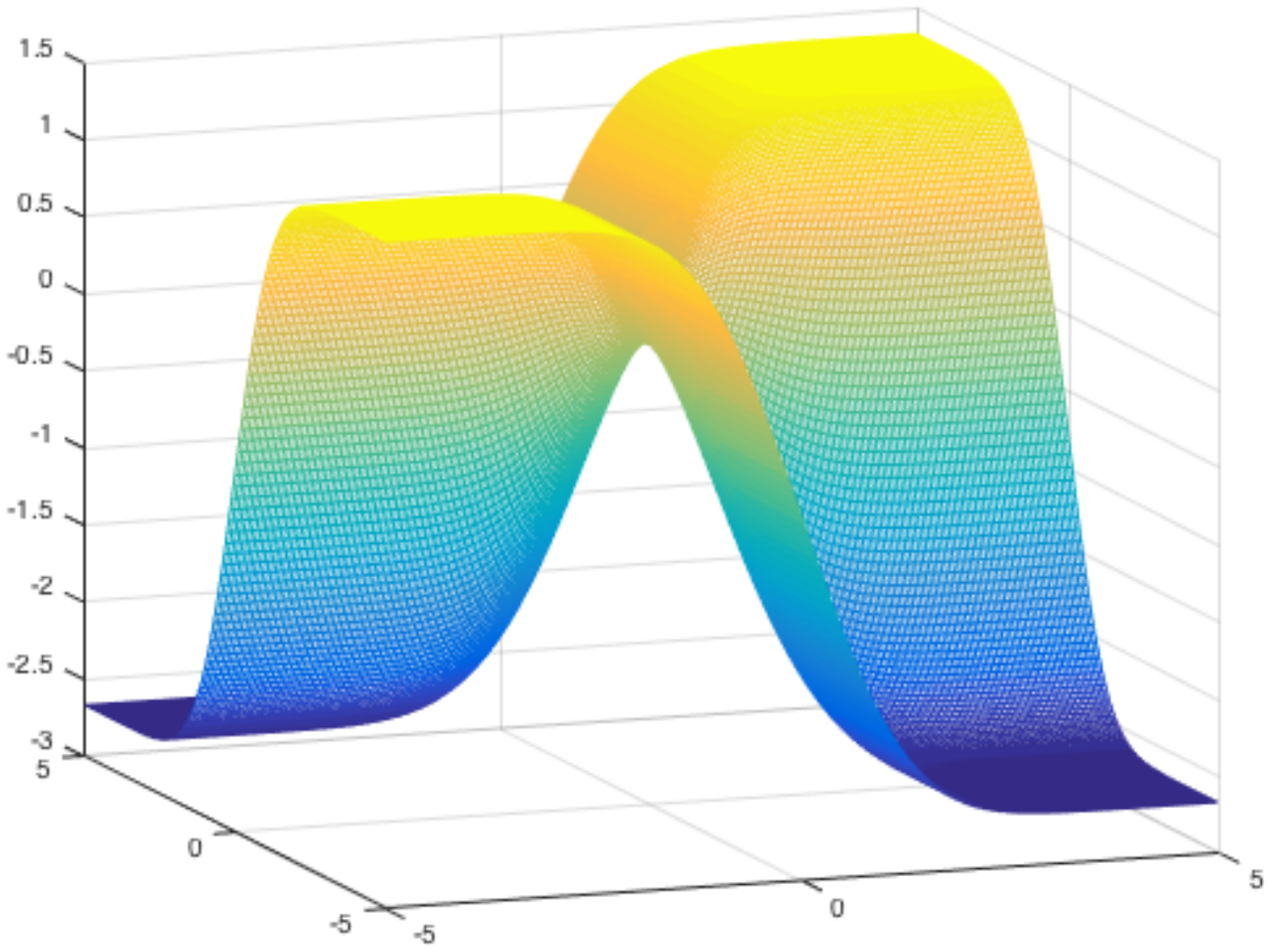} \vspace{-0.25in}
\end{tabular}
\caption{Influence functions of correlation correlations $\rho_p, \rho_g$ and $\tau$  for the bivariate normal distribution with $\mu_x =\mu_y=0$, $\sigma_x =\sigma_y=1$ and $\rho =0.5$.}
\label{fig:IF}
\end{figure*}
Figure \ref{fig:IF} displays the influence function of each correlation coefficient for the bivariate normal distribution with $\mu_X =\mu_Y=0$, $\sigma_X =\sigma_Y=1$ and $\rho =0.5$.  Note that scales of the value of the influence functions in the three plots are quite different. 

\section{Estimation}
Let $\mb z_i=(x_i, y_i)^T$, and $\mathcal{Z}=(\mb z_1, \mb z_2,..., \mb z_n)$ be a random sample from a continuous distribution $H$ with an  empirical distribution $H_n$. Replacing $H$ in (\ref{grhod})  with $H_n,$  we have the sample counterpart of the symmetric Gini correlation coefficient $\rho_g(H_n)=\hat{\rho}_{g}$:
\begin{align*}
&\hat{\rho}_{g}=\displaystyle\frac{\sum_{1 \leq i<j \leq n}\frac{(x_i-x_j)(y_i-y_j)}{\sqrt{(x_i-x_j)^{2}+(y_i-y_j)^2}}}{\sqrt{\sum_{1 \leq i<j \leq n}\frac{(x_i-x_j)^2}{\sqrt{(x_i-x_j)^{2}+(y_i-y_j)^2}}} \sqrt{\sum_{1 \leq i<j \leq n}\frac{(y_i-y_j)^2}{\sqrt{(x_i-x_j)^{2}+(y_i-y_j)^2}}}}.
\end{align*}

Using the same notations in Section \ref{influenceF}, we have the following central limit theorem of the sample symmetric Gini correlation $\hat{\rho}_g$. 
\begin{theorem}{}{}\label{deltamethod}
Let $\mb z_1, \mb z_2,..., \mb z_n$ be a random sample from 2-dimensional distribution $H$ with finite second moment. Then $\hat{\rho}_{g}$ is an unbiased, $\sqrt{n}$-consistent estimator of $\rho_{g}.$ Furthermore,
$\sqrt{n}(\hat{\rho}_g -\rho_g) \stackrel{d}{\rightarrow} N(0, v_g)$ as $n \to \infty$,
where 
\begin{align*}
v_g&= \E[ \mbox{IF}((X,Y)^T,\rho_g, H)]^2=\frac{\rho^2_g}{4}\bigg(\frac{1}{T_1^2}\E[L_1^2(X,Y)]+\frac{4}{T_2^2}\E[ L_2^2(X,Y)]\\ 
&+\frac{1}{T_3^2}\E[ L_3^2(X,Y)]-\frac{4}{T_1T_2}\E L_1(X,Y)L_2(X,Y)+\frac{2}{T_1T_3}\E L_1(X,Y)L_3(X,Y)\\ 
&-\frac{4}{T_2T_3}\E L_2(X,Y)L_3(X,Y)\bigg).
\end{align*}
\end{theorem}
Although (\ref{ST2}) implies  Theorem \ref{deltamethod}, it is hard to check  regularity conditions for the von Mises expansion (\ref{ST}). Instead, we prove it in the Appendix using the multivariate delta method and the asymptotic normality of the sample Gini covariance matrix, which is based on the $U$-statistics theory (***, 2015).

For an elliptical distribution $H$, Theorem 2 shows that  $\hat{\rho}_g$ is not a Fisher consistent estimator of $\rho$. We need to consider the inverse transformation $\hat{\rho}^g=k^{-1}(\hat \rho_g)$, where the function $k$ is given in (\ref{relation}).   Applying the delta method,  we obtain the $\sqrt{n}$-consistency of estimator $\hat{\rho}^g$ for $\rho$. 
\begin{theorem}{}{}\label{consistentgini}
Let $\mb z_1,\mb z_2,...,\mb z_n$ be a sample from elliptical distribution $H$ with finite second moment and $\mb\Sigma=\sigma^2\begin{pmatrix}1 \;\;\;&\rho\\ \rho\;\;\;&1\end{pmatrix}$. Then $\hat{\rho}^g = k^{-1}(\hat{\rho}_g)$ is unbiased and a $\sqrt{n}$-consistent estimator of $\rho$. Moreover,
$\sqrt{n}(\hat{\rho}^g-\rho) \stackrel{d}{\rightarrow} N(0, [1/k'(\rho)]^2 v_g)$ as $n \to \infty$,
where the function $k$ is given in (\ref{relation}), $v_g$ is given in Theorem \ref{deltamethod}, and $k'(\rho)$ is
\begin{align*}
&k'(\rho)=
&\frac{-3 (\rho+1) \mbox{EE}^{2}(\frac{2 \rho}{\rho+1})+4\mbox{EE}(\frac{2 \rho}
{\rho+1})\mbox{EK}(\frac{2 \rho}{\rho+1})+(\rho-1) \mbox{EK}^{2}(\frac{2 \rho}
{\rho+1}))}{2(\rho+1)\rho^{2}\mbox{EE}^{2}(\frac{2 \rho}{\rho+1})}.
\end{align*} 
\end{theorem}

Theorem \ref{consistentgini} provides an estimator based on $\hat{\rho}_g$ for the correlation parameter for  elliptical distributions. The asymptotic variance $[k'(\rho)]^{-2}v_g$ can be used to evaluate asymptotic efficiency of $\hat{\rho}^g$.

\subsection{Asymptotic efficiency}
  To compare relative efficiency, we present the asymptotic variances (ASV) of four estimators of $\rho$ including Pearson's estimator $\hat{\rho}_p$, $\hat{\rho}^g$, the regular Gini correlation estimator, and the estimator through Kendall's $\tau$ estimator. 
  
Witting and M\"{u}ller-Funk (1995) established asymptotic normality for the regular sample Pearson correlation coefficient $\hat{\rho}_p$:
\begin{align*}
\sqrt{n}(\hat{\rho}_p-\rho) \stackrel{d}{\rightarrow} N(0, v_p)\;\; as\;\; n \to \infty,  
\end{align*}
where
\begin{align*}
v_p=(1+\frac{\rho^2}{2})\frac{\sigma_{22}}{\sigma_{20}\sigma_{02}}+\frac{\rho^2}{4}(\frac{\sigma_{40}}{\sigma^2_{20}}+\frac{\sigma_{04}}{\sigma^2_{02}}-\frac{4 \sigma_{31}}{\sigma_{11}\sigma_{20}}-\frac{4 \sigma_{13}}{\sigma_{11}\sigma_{02}}), 
\end{align*}
and $\sigma_{kl}=\E[(X-\E X)^{k}(Y-\E Y)^l]$. The Pearson correlation estimator requires a finite fourth moment on the distribution to evaluate its asymptotic variance.
For bivariate normal distributions, the asymptotic variance $v_p$ simplifies to $(1-\rho^2)^2$.

An estimator $\hat{\rho}_{\gamma}$ of the regular Gini correlation $\gamma(X,Y)$ is 
\begin{align*}
\hat{\rho}_{\gamma}= \frac{ {n \choose 2}^{-1}\sum_{1\le i<j \le n} h_1(\mb{z}_i, \mb{z}_j)}{ {n \choose 2}^{-1}\sum_{1\le i<j\le n}h_2( \mb{z}_i, \mb{z}_j)}, 
\end{align*}
where 
$$h_{1}( \mb{ z}_1, \mb{z}_2)=[(x_1-x_2)I (y_1>y_2)+(x_2-x_1)I(y_2>y_1)]/4$$ and $h_2(\mb{ z}_1, \mb{ z}_2)=|x_1-x_2|/4.$
Using U-statistic theory, Schechtman and Yitzhak (1987) provided the asymptotic normality:
\begin{align*}
\sqrt{n}(\hat{\rho}_{\gamma}-\rho_{\gamma})\stackrel{d} \rightarrow N(0, v_{\gamma}) \;\;\;as \;\;\; n\rightarrow \infty,
\end{align*}
with  
\begin{align*}
v_{\gamma}=({4}/{\theta^2_2}) \zeta_1(\theta_1)+({4\theta^2_{1}}/{\theta^4_2})\zeta_2(\theta_2)-({8\theta_{1}}/{\theta^3_2}) \zeta_3(\theta_1,\theta_2), 
\end{align*}
where $$\theta_1=\mbox{cov}(X, G(Y)), \;\;\theta_2=\mbox{cov}(X, F(X)),$$
$$\zeta_1(\theta_1)=\E_{\mb {z}_1}\left\{ \E_{\mb{ z}_2} [h_1(\mb{Z}_1,\mb{Z}_2)]\right\}^2-\theta_1^2,$$ 
$$\zeta_2(\theta_2)=\E_{\mb {z}_1}\left \{\E_{\mb{ z}_2}[h_2(\mb{ Z}_1,\mb{ Z}_2)]\right\}^2-\theta_2^2 $$ and 
$$\zeta_3(\theta_1, \theta_2)=\E_{\mb{ z}_1}\left \{\E_{\mb{ z}_2}[h_1(\mb{ Z}_1,\mb{ Z}_2)]  \E_{\mb{ z}_2}[h_2(\mb{ Z}_1,\mb{ Z}_2)]\right\}-\theta_1\theta_2.$$ Under elliptical distributions, $\gamma(X,Y)=\gamma(Y,X)=\rho$, hence the asymptotic variance of $\hat{\rho}_{\gamma}$ is $v_{\gamma}$.  For a normal distribution, Xu {\em et al.} (2010) provided an explicit formula of $v_{\gamma}$, given by $v_{\gamma} = \pi/3 +(\pi/3+4\sqrt{3})\rho^2-4\rho \arcsin (\rho/2)-4\rho^2\sqrt{4-\rho^2}$.

Borovskikh (1996) presented the asymptotic normality of the estimator $\hat{\tau}$: 
\begin{align*}
\sqrt{n}(\hat{\tau}-\tau) \stackrel{d}{\rightarrow} N(0, v_{\tau}) \;\; as\;\; n \to \infty,
\end{align*}
with
\begin{align*}
v_{\tau} =4 \E\{\E_{\mb z_1}^2(\mbox{sgn}[(X_2-X_1)(Y_2-Y_1)])\}- 4\E^2\{\mbox{sgn}[(X_2-X_1)(Y_2-Y_1)]\}.
\end{align*}
Applying the delta method to $\hat{\rho}_{\tau}=\sin(\pi \hat{\tau}/2)$, we obtain the asymptotic variance of $\hat{\rho}_{\tau}$ to be
$\frac{\pi^2}{4}(1-\rho^2) v_{\tau}$.  
Under a normal distribution, the asymptotic variance of $\hat{\rho}_{\tau}$   is  $\pi^2(1-\rho^2)(\frac{1}{9}-\frac{4}{\pi^2}\arcsin^2(\frac{\rho}{2}))$ (Croux and Dehon, 2010).

We compare asymptotic efficiency of the four estimators $\hat{\rho}^g$, $\hat{\rho}_{\gamma}$, $\hat{\rho}_{\tau}$ and $\hat{\rho}_p$ under three bivariate elliptical distributions (\ref{pdf}) with different fatness on the tail regions: the normal distributions with $g(t)=\frac{1}{2\pi}e^{-t/2}$; the $t$-distributions with $g(t)=\frac{1}{2\pi}(1+t/\nu)^{-\nu/2-1},$ where $\nu$ is the degree of freedom;  and the Kotz type distribution with $g(t)=\frac{1}{2\pi}e^{-\sqrt{t}}$. The normal distribution is the limiting distribution of the $t$-distributions as $\nu\rightarrow \infty$. The Kotz type distribution is a bivariate generalization of the Laplace distribution with the tail region fatness between that of the normal and $t$ distributions (Fang, Kotz and Hg, 1987). We consider only elliptical distributions because all four estimators $\hat{\rho}^g$, $\hat{\rho}_{\gamma}$, $\hat{\rho}_{\tau}$ and $\hat{\rho}_p$ are Fisher consistent for parameter $\rho$. The estimators for non-elliptical distributions may estimate different quantities, resulting in their asymptotical variances incomparable. 
\begin{table}[thb]
\center
\caption{Asymptotic relative efficiencies (ARE) of estimators $\hat{\rho}^g$, $\hat{\rho}_{\gamma}$ and $\hat{\rho}_{\tau}$ relative to $\hat{\rho}_p$ for different distributions, with asymptotic variance (ASV($\hat{\rho}_p$)) of Pearson estimator $\hat{\rho}_p$. }
\label{tab:ARE}
\begin{tabular}{||lc  || c|c|c||c ||}
\hline  \hline
 \multicolumn{2}{||c||}{Distribution} &\multicolumn{ 1}{c|}{ARE($\hat{\rho}^g, \hat{\rho}_p$)}  &\multicolumn{ 1}{c|}{ARE($\hat{\rho}_{\gamma}, \hat{\rho}_p$)}   &\multicolumn{ 1}{c||}{ARE($\hat{\rho}_{\tau},\hat{\rho}_p$) }   &\multicolumn{ 1}{c||}{ASV($\hat{\rho}_p$)} \\
\hline\hline
      & $\rho=0.1$ &  0.9321 &0.9558& 0.9125  &0.9816 \\
Normal & $\rho=0.5$ & 0.9769&0.9398& 0.8925 &0.5631 \\
&      $\rho=0.9$ & 0.9601 & 0.9004&  0.8439 &0.0361  \\
\hline
 & $\rho=0.1$ & 1.0182 &1.0304& 1.0146 &1.1558 \\
$t(15)$ & $\rho=0.5$ &1.0560& 0.9852& 0.9896  & 0.6643  \\
& $\rho=0.9$ &  1.0289&0.9468&  0.8804& 0.0427   \\
\hline
 & $\rho=0.1$ & 2.0095 &1.9502& 2.2586 &2.8800  \\
$t(5)$ &  $\rho=0.5$ & 1.9795&1.7666&  2.1060  &1.5961  \\
&  $\rho=0.9$ &  1.8629&1.5346& 1.7940 &0.1019  \\
\hline
 &$\rho=0.1$ &  1.2081 &1.1385 &1.2171& 1.6382  \\
Kotz & $\rho=0.5$ &  1.1850  &1.0854& 1.1510& 0.9378  \\
&  $\rho=0.9$ & 1.1599  & 0.9789 &1.0256 &0.0602  \\
\hline\hline
\end{tabular}
\end{table}

Without loss of generality, we consider only cases with $\rho>0$. Listed in Table \ref{tab:ARE} are asymptotic variances (ASV) of Pearson estimator $\hat{\rho}_p$, and asymptotic relative efficiencies (ARE) of estimators $\hat{\rho}^g$, $\hat{\rho}_{\gamma}$ and  $\hat{\rho}_{\tau}$ relative to $\hat{\rho}_p$ for different elliptical distributions under the homogeneous assumption,  where the asymptotic relative efficiency of an estimator with respect to another is defined as 
$\text{ARE}(\hat{\rho}_1, \hat{\rho}_2) = {\text{ASV}(\hat{\rho}_2)}/{\text{ASV}(\hat{\rho}_1)}$. 
The asymptotic variance of each estimator is obtained based on a combination of numeric integration and the Monte Carlo simulation. 

Table 1 shows that the asymptotic variances of $\hat{\rho}_p$, $\hat{\rho}^g$, $\hat{\rho}_{\gamma}$ and $\hat{\rho}_{\tau}$ all decrease as $\rho$ increases. When $\rho =1$, every estimator is equal to 1 without any estimation error. Asymptotic variances increase for $t$ distributions as the degrees of freedom $\nu$ decreases.  Under normal distributions, the Pearson correlation estimator is the maximum likelihood estimator of $\rho$, thus is most efficient asymptotically. The symmetric Gini estimator $\hat{\rho}^g$ is high in efficiency with  ARE's greater than 93\%; it is more efficient than Kendall's estimator $\hat{\rho}_\tau$. For heavy-tailed distributions, the symmetric Gini estimator is more efficient than Pearson's estimator $\hat{\rho}_p$. The AREs of the symmetric Gini estimator are close to those of Kendall's estimator $\hat{\rho}_\tau$ for Kotz samples.
Comparing with the regular Gini correlation estimator, the proposed measure has higher efficiency for all cases except for $\rho=0.1$ under normal and $t(15)$ distributions, in which the efficiency is about $2.4\%$ and $1.2\%$ lower.  
These results may be explained by  that the joint spatial rank used in $\hat{\rho}^g$ takes more dependence information than the marginal rank used in $\hat{\rho}_{\gamma}$. 

In summary, the proposed symmetric Gini estimator has nice asymptotic behavior that well balances between efficiency and robustness. It  is more efficient than the regular Gini, which  is also symmetric under elliptical distributions.


\subsection{Finite sample efficiency}\label{fse}
We conduct a small simulation to study the finite sample efficiencies of the correlation estimators for the symmetric Gini, regular Gini, Kendall's $\tau$ and Pearson correlations. $M=3000$ samples of two different sample sizes, $n=30, 300$, are drawn from $t$-distributions with $1, 3, 5, 15$ and $\infty$ degrees of freedoms and from the Kotz distribution. We use R Package ``mnormt" to generate samples from multivariate $t$ and normal  distributions (referred as $t(\infty)$ in  Table \ref{tab:2}).  For the Kotz sample, we first generate uniformly distributed random vectors on the unit circle by $\mb u=(\cos \theta, \sin\theta)^T$ with $\theta$ in $[0,2\pi]$, then generate $r$ from a Gamma distribution with $\alpha=2$ (the shape parameter) and $\beta=1$ (the scale parameter) and hence $\mb\Sigma^{1/2} r\mb u +\mb{\mu}$ is a sample from bivariate Kotz$(\mb\mu, \mb\Sigma)$. For more details, see *** (2015).

For each sample $m$, each estimator $\hat{\rho}^{(m)}$ is calculated and the root of mean squared error (RMSE)  of the estimator $\hat{\rho}$ is computed as
\begin{align*}
\mbox{RMSE}(\hat{\rho})=\sqrt{\frac{1}{M}\sum_{m=1}^{M}(\hat{\rho}^{(m)}-\rho)^2}.
\end{align*}
The procedure is repeated 100 times. In Table \ref{tab:2}, we report the mean and standard deviation (in parentheses) of $\sqrt{n}$RMSEs of correlation estimators $\hat{\rho}^g$, $\hat{\rho}_{\gamma}$, $\hat{\rho}_{\tau}$ and $\hat{\rho}_p$ when the scatter matrix is homogeneous with $\mb{\Sigma}=\sigma^2\begin{pmatrix}1 \;\;\;&\rho\\\rho\;\;\;&1 \end{pmatrix}$. The case of $n=\infty$ corresponds to the asymptotic standard deviation of each estimator that can be obtained from Table \ref{tab:ARE}. Since $\hat{\rho}^g$ cannot be given explicitly due to the inverse transformation involved in $\hat{\rho}^g = k^{-1}(\hat{\rho}_g)$, we use a numerical way to obtain $\hat{\rho}^g$ by creating a correspondence between $s$ and $t$, where $s=k(t)$ and $t$ is a very fine grid on $[0,1]$.  $\hat{\rho}_g$ is computed by using R package ``ICSNP" for spatial.rank function. 
\begin{table}
\caption{The mean and standard deviation (in parentheses) of $\sqrt{n}$RMSE of $\hat{\rho}^g$,  $\hat{\rho}_{\gamma}$, $\hat{\rho}_{\tau}$ and $\hat{\rho}_p$ under different distributions when the scatter matrix is homogeneous. }
\label{tab:2}
\footnotesize
\begin{tabular}{ lll | c |c |  c |  c}
\hline  \hline
 {Dist} &$\rho$&$n$             &{$\sqrt{n}$RMSE($\hat{\rho}^g$)}   & $\sqrt{n}$RMSE($\hat{\rho}_{\gamma}$)   &{$\sqrt{n}$RMSE($\hat{\rho}_{\tau}$)}   &{$\sqrt{n}$RMSE($\hat{\rho}_p$)}  \\ \hline \hline
& $\rho=0.1$ & $n=30$ &0.7767 (.0115) &1.0418 (.0115) &1.0785 (.0120) &1.0095 (.0120)\\
&          & ${\it n=300}$  & {\it 0.9648 (.0104)} &{\it 1.0184 (.0121)} &{\it 1.0427 (.0139)} &{\it 0.9925 (.0121)}\\[1ex]
$t(\infty)$& $\rho=0.5$& $n=30$ & 0.7887 (.0110) &0.8150 (.0115) &0.8517 (.0126) &0.7827 (.0115)\\
   &        & ${\it n=300}$  & {\it 0.7638 (.0087)} &{\it 0.7777 (.0104)} &{\it 0.8002 (.0104)} &{\it 0.7534 (.0104)}\\  [1ex]
&$\rho=0.9$   & $n=30$ &0.2147 (.0044) &0.2306 (.0044) &0.2541 (.0049) &0.2103 (.0044)\\ 
  &         & ${\it n=300}$   &{\it 0.1957 (.0017)} &{\it 0.2026 (.0035)} &{\it 0.2113 (.0035)} &{\it 0.1923 (.0017)}\\  \hline

&$\rho=0.1$ & $n=30$ & 0.8013 (.0120) &1.0828 (.0120) &1.1026 (.0115) &1.0735 (.0115)\\
 &          & ${\it n=300}$   & {\it 1.0011 (.0104)} &{\it 1.0669 (.0121)} &{\it 1.0721 (.0139)} &{\it 1.0756 (.0121)}\\[1ex]
$t(15)$& $\rho=0.5$&$n=30$ & 0.8177 (.0115)& 0.8506 (.0126)& 0.8731 (.0131)& 0.8347 (.0126)\\
   &        & ${\it n=300}$    & {\it 0.7985 (.0104)}&{\it 0.8227 (.0104)}& {\it 0.8279 (.0104)}& {\it 0.8193 (.0104)}\\ [1ex]
&$\rho=0.9$ & $n=30$&0.2251 (.0044) &0.2432 (.0044 )&0.2635 (.0164) &0.2262 (.0044)\\ 
  &         & ${\it n=300}$   & {\it 0.2044 (.0035) }&{\it 0.2165 (.0035)} &{\it 0.2200 (.0035)} &{\it 0.2078 (.0035)}\\  \hline

&$\rho=0.1$ & $n=30$ & 0.8698 (.0137)& 1.2083 (.0126)& 1.1562 (.0131)& 1.2987 (.0137)\\
&          & ${\it n=300}$   &{\it 1.1085 (.0121)} &{\it 1.2246 (.0156)} &{\it 1.1310 (.0139)} &{\it 1.5155 (.0242)}\\[1ex]
$t(5)$& $\rho=0.5$ & $n=30$ & 0.9032 (.0110) &0.9580 (.0126)& 0.9202 (.0126) &1.0221 (.0159)\\
   &        & ${\it n=300}$   & {\it 0.9007 (.0121)} &{\it 0.9492 (.0121)} &{\it 0.8764 (.0121)} &{\it 1.1535 (.0208)}\\ [1ex]
& $\rho=0.9$   & $n=30$ & 0.2569 (.0164) &0.2859 (.0066) &0.2832 (.0164) &0.2908 (.0088)\\ 
  &         & ${\it n=300}$   & {\it 0.2338 (.0069)} &{\it 0.2615 (.0035)} &{\it 0.2408 (.0069)} &{\it 0.2996 (.0087)}\\  \hline

&$\rho=0.1$  & $n=30$ & 0.9706 (.0137)& 1.3923 (.0170) &1.2050 (.0142) &1.6459 (.0214)\\
&      & ${\it n=300}$  &{\it  1.2921 (.0156)} &{\it 1.5329 (.0191)} &{\it 1.1865 (.0156)} &{\it 2.7782 (.0554)}\\[1ex]
$t(3)$& $\rho=0.5$& $n=30$ &1.0231 (.0131) &1.1201 (.0170) &0.9651 (.0148) &1.3343 (.0246)\\
   &        & ${\it n=300}$   &{\it 1.1068 (.0173)}& {\it 1.2142 (.0208)}& {\it 0.9284 (.0121)}& {\it 2.1876 (.0675)}\\ [1ex]
&$\rho=0.9$ & $n=30$ &  0.3127 (.0104) &0.3642 (.0131) &0.3051 (.0066) &0.4289 (.0236)\\ 
  &         & ${\it n=300}$   &{\it  0.2944 (.0104)} &{\it 0.3672 (.0173)} &{\it 0.2615 (.0035)}&{\it 0.6564 (.0658)}\\  \hline

&$\rho=0.1$  & $n=30$ & 1.7418 (.0301) &2.7222 (.0285 )&1.3704 (.0170) &3.3104 (.0279)\\
&      & ${\it n=300}$  &{\it  4.3423  (.0814)}  &{\it 6.7879  (.0918)}  &{\it 1.3735  (.0173)} &{\it 10.256  (.0918)}\\[1ex]
$t(1)$& $\rho=0.5$& $n=30$ &1.6706 (.0153) &2.3892 (.0361) &1.1184 (.0164) &2.9687 (.0466)\\
   &        & ${\it n=300}$   & {\it 4.2574 (.0485)} &{\it 5.9357 (.1057)} &{\it 1.0999 (.0156)} &{\it 9.1781 (.1472)}\\ [1ex]
&$\rho=0.9$ & $n=30$ & 0.9065 (.0361) &1.2083 (.0586) &0.4004 (.0088) &1.5917 (.0728)\\ 
  &         & ${\it n=300}$   & {\it 2.1616 (.1074)} &{\it 2.9947 (.1784)} &{\it  0.3464 (.0052)} &{\it 4.9589 (.2182)}\\  \hline
& $\rho=0.1$& $n=30$ & 0.8692 (.0126) &1.2083 (.0148) &1.1842 (.0148) &1.2389 (.0148)\\
&          & ${\it n=300}$   & {\it 1.0947 (.0139)} &{\it 1.2055 (.0173)} &{\it 1.1639 (.0156)} &{\it 1.2713 (.0173)}\\[1ex]

Kotz& $\rho=0.5$ & $n=30$ &0.9037 (.0137) &0.9569 (.0148) &0.9465 (.0142) &0.9711 (.0170)\\
   &        & ${\it n=300}$  & {\it 0.8903 (.0121)} &{\it 0.9318 (.0121)} &{\it 0.9059 (.0121)} &{\it 0.9665 (.0121)}\\ [1ex]

& $\rho=0.9$   & $n=30$ & 0.2563 (.0164) &0.2832 (.0164) &0.2952 (.0060)& 0.2706 (.0060)\\ 
  &         & ${\it n=300}$   & {\it 0.2304 (.0035)}& {\it 0.2529 (.0035)}& {\it 0.2494 (.0035)} &{\it 0.2477 (.0035)}\\  \hline\hline
\end{tabular}
\end{table}

In Table \ref{tab:2},  the $\sqrt{n}$RMSEs demonstrate an increasing trend as $\rho$ decreases or as the degree of freedom $\nu$ decreases for $t$ distributions.
For $n=300$, the behavior of each estimator is similar to its asymptotic efficiency behavior.  For example, for $n=300$ and $\rho=0.5$ under the normal distribution, the $\sqrt{n}$RMSE of $\hat{\rho}_p$ is 0.7534 close to the asymptotic standard deviation 0.7504. We include heavy-tailed $t(1)$ and $t(3)$ distributions in the simulation to demonstrate finite sample behavior of Pearson and Gini estimators when their asymptotic variances may not exist. $\sqrt{n}$RMSE of $\hat{\rho}_p$ is about twice as that of $\hat{\rho}^g$ for $n=300$ in both $t(1)$ and $t(3)$ distributions.  For $t(1)$ distribution, $\hat{\rho}_\tau$ is much better than others in terms of $\sqrt{n}$RMSE. When the sample size is small ($n=30$), $\hat{\rho}^g$ performs the best. The $\sqrt{n}$RMSEs of $\hat{\rho}^g$ are smaller than that of $\hat{\rho}_\tau$ even under heavy-tailed $t(3)$ distribution. $\hat{\rho}^g$ has a smaller $\sqrt{n}$RMSE than the Pearson correlation estimator for the normal distribution with $\rho=0.1$ and all other distributions. The symmetric Gini  estimator $\hat{\rho}^g$ has smaller $\sqrt{n}$RMSE than the regular Gini estimator $\hat{\rho}_\gamma$ for all cases we consider.  The simulation demonstrates superior finite sample behavior of the proposed estimator.



\section{The affine invariant version of symmetric Gini correlation}
The proposed  $\rho_g$ in Section \ref{sec:sgc} is only invariant under translation and homogeneous change. We now  provide an affine invariant version of $\rho_g$, denoted as $\rho_G$,  in order to gain the invariance property under heterogeneous changes. This is based on the affine equivariant (AE) Gini covariance matrix $\mb{\Sigma}_G$ proposed by *** (2015).

The basic idea of $\mb{\Sigma}_G$ is that the Gini covariance matrix on standardized data should be proportional to the identity matrix $\mb I$. That is,
$\E (\mb\Sigma_G^{-1/2} \mb Z) \mb r^T(\mb\Sigma_G^{-1/2}\mb Z) = c\mb I,$ where $c$ is a positive constant.
In other words, the AE version of the Gini covariance matrix is the solution of
\begin{align}\label{trgcm}
\E \frac{\mb\Sigma_G^{-1/2}(\mb Z_1-\mb Z_2)(\mb Z_1-\mb Z_2)^{T}\mb\Sigma_G^{-1/2}}{\sqrt{(\mb Z_1-\mb Z_2)^{T}\mb\Sigma_G^{-1}(\mb Z_1-\mb Z_2)}}=c(H)\mb I,
\end{align}
where $c(H)$ is a constant depending on $H$. In this way, the matrix valued functional $\mb\Sigma_{G}(\cdot)$ is a scatter matrix in the sense that for any nonsingular matrix $A$ and vector $\mb b$, $\mb\Sigma_{G}(A \mb Z +\mb b)=A \mb\Sigma_{G}(\mb Z)A^{T}.$

Let $\mb Z =(X,Y)^T$ be  a bivariate random vector with  distribution function $H$ and $\mb\Sigma_{G}:=\begin{pmatrix} G_{11}& G_{12}\\ G_{21}&G_{22} \end{pmatrix}$ be the solution of (\ref{trgcm}). Then the affine invariant version of $\rho_g$  is defined as
$\rho_{G}(X,Y)=\frac{G_{21}}{\sqrt{G_{11}}\sqrt{G_{22}}}$.
Since the value of $c(H)$ in (\ref{trgcm}) does not change the value of $\rho_G(X,Y)$, without loss of generality, assume $c(H)=1$.
\begin{table}
\caption{The mean and standard deviation (in parentheses) of  $\sqrt{n}$RMSE of $\hat{\rho}_G$, $\hat{\rho}_{\gamma}$,  $\hat{\rho}_{\tau}$ and $\hat{\rho}_p$ under different distributions with a heterogeneous scatter matrix. }
\label{tab:3}
\footnotesize
\begin{tabular}{ lll | c |c |  c |  c}
\hline  \hline
 {Dist} &$\rho$&$n$ &{$\sqrt{n}$RMSE($\hat{\rho}_G$)}   & $\sqrt{n}$RMSE($\hat{\rho}_{\gamma}$)   &{$\sqrt{n}$RMSE($\hat{\rho}_{\tau}$)}   &{$\sqrt{n}$RMSE($\hat{\rho}_p$)}  \\ \hline \hline
 &
$\rho=0.1$& $n=30$ & 1.0171 (.0126)&1.0401 (.0126)&1.0768 (.0131)&1.0073 (.0120)\\
   &        & ${\it n=300}$  & {\it 1.0011 (.0139)}&{\it 1.0133 (.0139)}&{\it 1.0392 (.0156)}&{\it 0.9890 (.0139)}\\ 
$t(\infty)$ &$\rho=0.5$  & $n=30$ & 0.7887 (.0120)&0.8123 (.0126)&0.8501 (.0137)&0.7800 (.0120)\\
   &        & ${\it n=300}$  &{\it 0.7621 (.0104)}&{\it 0.7794 (.0104)}&{\it 0.8002 (.0104)}&{\it 0.7534 (.0104)}\\
& $\rho=0.9$ & $n=30$ & 0.2125 (.0022)&0.2306 (.0044)&0.2541 (.0049)&0.2098 (.0044)\\ 
  &         & ${\it n=300}$   & {\it 0.1940 (.0035)}&{\it 0.2026 (.0035)}&{\it 0.2113 (.0035)}&{\it 0.1923 (.0017)}\\  \hline

&$\rho=0.1$  & $n=30$ &1.0582 (.0126)&1.0839 (.0131)&1.1042 (.0126)&1.0741 (.0126)\\
 &          & ${\it n=300}$   &{\it 1.0496 (.0121)}&{\it 1.0687 (.0121)}&{\it 1.0739 (.0121)}&{\it 1.0756 (.0121)}\\
$t(15)$& $\rho=0.5$ & $n=30$ & 0.8221 (.0099)&0.8506 (.0099)&0.8731 (.0110)&0.8353 (.0099)\\
   &        & ${\it n=300}$    & {\it 0.7967 (.0104)}&{\it 0.8210 (.0121)}&{\it 0.8279 (.0104)}&{\it 0.8175 (.0121)}\\
   
&$\rho=0.9$   & $n=30$ & 0.2224 (.0049)&0.2437 (.0049)&0.2635 (.0060)&0.2262 (.0049)\\
  &    & ${\it n=300}$   & {\it 0.2026 (.0035)}&{\it 0.2165 (.0035)}&{\it 0.2200 (.0035)}&{\it 0.2078 (.0035)}\\  \hline

& $\rho=0.1$ & $n=30$ & 1.1727 (.0164)&1.2072 (.0148)&1.1557 (.0153)&1.2981 (.0192)\\
 &          & ${\it n=300}$   & {\it 1.1847 (.0156)}&{\it 1.2246 (.0156)}&{\it 1.1310 (.0139)}&{\it 1.5155 (.0242)}\\
$t(5)$& $\rho=0.5$ & $n=30$ & 0.9169 (.0120)&0.9585 (.0115)&0.9213 (.0120)&1.0226 (.0137)\\
   &        & ${\it n=300}$  & {\it 0.8989 (.0139)}&{\it 0.9492 (.0139)}&{\it 0.8764 (.0121)}&{\it 1.1553 (.0242)}\\
   
&$\rho=0.9$ & $n=30$ & 0.2520 (.0060)&0.2865 (.0071)&0.2832 (.0060)&0.2919 (.0110)\\ 
  &         & ${\it n=300}$   & {\it 0.2304 (.0035)}&{\it 0.2615 (.0035)}&{\it 0.2408 (.0035)}&{\it 0.2979 (.0087)}\\  \hline

&$\rho=0.1$ & $n=30$ & 1.3540 (.0519)&1.3918 (.0159)&1.2039 (.0142)&1.6475 (.0203)\\
 &          & ${\it n=300}$   & {\it 1.4497 (.0225)}&{\it 1.5346 (.0225)}&{\it 1.1847 (.0156)}&{\it 2.7782 (.0606)}\\
$t(3)$& $\rho=0.5$  & $n=30$  &1.0670 (.0159)&1.1190 (.0170)&0.9629 (.0148)&1.3321 (.0219)\\
   &        & ${\it n=300}$    &{\it 1.1033 (.0139)}&{\it 1.2090 (.0173)}&{\it 0.9249 (.0121)}&{\it 2.1910 (.0606)}\\
&$\rho=0.9$   & $n=30$ & 0.3095 (.0099)&0.3681 (.0137)&0.3062 (.0066)&0.4376 (.0230)\\ 
  &         & ${\it n=300}$   & {\it 0.2841(.0069)}&{\it 0.3655 (.0156)}&{\it 0.2615 (.0035)}&{\it 0.6461 (.0675)}\\  \hline

&$\rho=0.1$ & $n=30$ &2.7622 (.0274)&2.7244 (.0268)&1.3693 (.0192)&3.3148 (.0268)\\
 &          & ${\it n=300}$   &{\it 6.8381 (.0970)}&{\it 6.7879 (.0797)}&{\it 1.3770 (.0173)}&{\it 10.259 (.0901)}\\
$t(1)$& $\rho=0.5$  & $n=30$  & 2.4133 (.0433)&2.3831 (.0372)&1.1206 (.0164)&2.9643 (.0466)\\
   &        & ${\it n=300}$    &{\it 5.8768 (.1386)}&{\it 5.9132 (.1178)}&{\it 1.0947 (.0139)}&{\it 9.1522 (.1455)}\\
&$\rho=0.9$   & $n=30$ & 1.1875 (.0608)&1.2148 (.0537)&0.4009 (.0088)&1.6015 (.0635)\\ 
  &         & ${\it n=300}$   &{\it 2.7747 (.2148)}&{\it 2.9930 (.1853)}&{\it 0.3481 (.0052)}&{\it 4.9727 (.2113)}\\  \hline

&$\rho=0.1$ & $n=30$ &1.1672 (.0131)&1.2066 (.0131)&1.1831 (.0142)&1.2368 (.0142)\\
 &          & ${\it n=300}$   & {\it 1.1674 (.0139)}&{\it 1.2038 (.0139)}&{\it 1.1605 (.0139)}&{\it 1.2731 (.0156)}\\
Kotz& $\rho=0.5$  & $n=30$ & 0.9136 (.0148)&0.9574 (.0148)&0.9454 (.0153)&0.9706 (.0148)\\
   &        & ${\it n=300}$   &{\it 0.8885 (.0121)}&{\it 0.9336 (.0121)}&{\it 0.9059 (.0121)}&{\it 0.9665 (.0121)}\\ 
& $\rho=0.9$  & $n=30$  & 0.2503 (.0049)&0.2815 (.0060)&0.2941 (.0060)&0.2684 (.0055)\\ 
  &         & ${\it n=300}$   &{\it  0.2269 (.0035)}&{\it 0.2546 (.0035)}&{\it 0.2511 (.0035)}&{\it 0.2477 (.0035)}\\  \hline\hline

\end{tabular}
\end{table}

\begin{theorem}{}{}\label{affine}
For any bivariate random vector $\mb Z=(X, Y)^T$ having an elliptical distribution $H$ with finite first moment,  $\rho_{G}(aX, bY)=sgn(ab)\rho_{G}(X, Y)$ for any $ab\neq 0$.
\end{theorem}
\begin{remark}{}{}
Under elliptical distributions, 
$\rho_G=\rho$. This is true since $\mb\Sigma_G =\mb\Sigma$ for elliptical distributions.
\end{remark}

When  a random sample $\mb z_1, \mb z_2, ..., \mb z_n $ is available, replacing $H$ with its empirical distribution $H_n$ in (\ref{trgcm}) yields the sample counterpart $\hat{\mb\Sigma}_G$, and hence the sample $\hat{\rho}_G$ is obtained accordingly.  We  obtain $\hat{\mb\Sigma}_G$ by a common re-weighted iterative algorithm:
\begin{equation*}
\hat{\mb\Sigma}_G^{(t+1)} \longleftarrow  \frac{2}{n(n-1)}\sum_{1\leq i <j \leq n}\frac{(\mb z_i-\mb z_j)(\mb z_i-\mb z_j)^T}{\sqrt{(\mb z_i-\mb z_j)^T (\hat{\mb\Sigma}_G^{(t)})^{-1}(\mb z_i-\mb z_j)}}.
\end{equation*}
The initial value can take $\hat{\mb\Sigma}_G^{(0)}=\mb I_d$. The iteration stops when $\|\hat{\mb\Sigma}_G^{(t+1)}-\hat{\mb\Sigma}_G^{(t)}\|<\varepsilon$ for a pre-specified number $\varepsilon>0$, where $\|\cdot\|$ can take any matrix norm.

Next, we study finite sample efficiency of $\hat{\rho}_G$ under the same simulation setting as in Section \ref{fse} except that the scatter matrix is heterogeneous. The scatter matrix of each elliptical distribution is $\mb{\Sigma}=\begin{pmatrix}1 \;\;&2\rho\\2\rho\;\;&4 \end{pmatrix}$. Table \ref{tab:3} reports $\sqrt{n}$RMSE of correlation estimators $\hat{\rho}_G$, $\hat{\rho}_{\gamma}$, $\hat{\rho}_{\tau}$ and $\hat{\rho}_p$.   The numbers in the last three columns are very close to those in Table \ref{tab:2} because $\hat{\rho}_{\gamma}$, $\hat{\rho}_{\tau}$ and $\hat{\rho}_p$ are affine invariant. $\sqrt{n}$RMSEs of $\hat{\rho}_G$ are also close to  $\sqrt{n}$RMSE of $\hat{\rho}^g$ for $n=300$, but are larger than those for $n=30$ and $\rho=0.1$.  The loss of finite sample efficiency of $\hat{\rho}_G$ for a small size under low dependence $\rho$ is probably caused by the iterative algorithm in the computation of $\hat{\rho}_G$. The problem is even worse in $t(1)$ distribution where the first moment does not exist.   As the value of $\rho$ increases, $\sqrt{n}$RMSE of each estimator decreases for all distributions. Under Kotz and $t(15)$ distributions,  the affine invariant Gini estimator $\hat{\rho}_G$ is the most efficient; under $t(5)$ distribution, the $\sqrt{n}$RMSE of $\hat{\rho}_G$ is smaller than that of Kendall's $\hat{\rho}_\tau$ when $\rho=0.9$.  For the normal distributions, $\hat{\rho}_G$ is almost as efficient as $\hat{\rho}_p$ when $\rho=0.9$. The affine invariant Gini correlation estimator shows a good finite sample efficiency. Again, the proposed Gini has smaller $\sqrt{n}$RMSEs than the regular Gini in all cases.

\section{Application}
For the purpose of illustration, we apply the symmetric Gini correlations to the famous Fisher's Iris data which is available in R. The data set consists of 50 samples from each of three species of Iris (Setosa, Versicolor and Virginica). Four features are measured in centimeters from each sample: sepal length (Sepal L.), sepal width (Sepal W.), petal length (Petal L.), and petal width (Petal W.). The mean and standard deviation of each of the variables for all data and each species data are listed in Table \ref{tab:4}. All the three species have similar sizes in sepals. But Setosa has a much smaller petal size than the other two species.  Hence we shall study the correlation of the variables for each Iris species. 

\begin{table}[thb]
\caption{Summary Statistics of Variables in Iris Data}
\centering
\label{tab:4}
\begin{tabular}{l||c|ccc||c|ccc}\hline\hline
&\multicolumn{4}{c||}{Mean}&\multicolumn{4}{c}{Standard  Deviation}\\ \cline{2-9}
& All &Setosa&Vesicolor&Virginica&All &Setosa&Vesicolor&Virginica\\  \hline
Sepal L.& 5.843& 5.006 & 5.936&6.588&0.828&0.352&0.516&0.636\\  \hline
Sepal W.&3.057&3.428&2.770&2.974&0.436&0.379&0.314&0.322\\ \hline
Petal L. &3.758&1.462&4.260&5.552&1.765&0.174&0.470&0.552\\ \hline
Petal W. & 1.199&0.246&1.326&2.026& 0.762&0.105&0.198&0.275 \\ \hline\hline
\end{tabular}
\end{table}

For each Iris species,  we compute different correlation measures for all pairs of variables. Since variations of variables are quite different, the affine equivariant version of symmetric gini correlation estimator $\hat{\rho}_G$ is used. For each pair of variables $X$ and $Y$, we also calculate Pearson correlation, Kendall's $\tau$ and two regular gini correlation estimators, denoted as $\hat{\gamma}_{1,2}$ ($\hat{\gamma}(X, Y)$) and  $\hat{\gamma}_{2,1}$ ($\hat{\gamma}(Y, X)$). All correlation estimators are listed in Table \ref{tab:5}. 

\renewcommand{\baselinestretch}{1}
\begin{table}[thb]
\caption{Pearson correlation, Kendal's $\tau$, Affine equivariant symmetric Gini correlation and  Regular Gini correlations of variables for Iris data set.}
\label{tab:5}
\footnotesize
\begin{tabular}{lc|c|c|c|c|c|c }
\hline  \hline

 &&  Sepal  L. &  Sepal L.  &  Sepal L. &  Sepal W.&  Sepal W. &  Petal L. \\
 {Species} &Correlations&\&&\&& \&&\&&\&&\& \\
&       & Sepal W. &  Petal L.  & Petal W. &  Petal L.&  Petal W.&  Petal W.  \\
\hline\hline

 &  $\hat{\rho}_p$ & 0.743 &0.267& 0.278 &0.178&0.233&0.332 \\
 &  $\hat{\tau}$& 0.597&0.217& 0.231 &0.143&0.234&0.222 \\
Setosa&  $\hat{\rho}_G$  & 0.742 &0.274& 0.285 &0.182&0.256&0.312 \\
&  $\hat{\gamma}_{1,2}$  & 0.759 &0.283& 0.261 &0.211&0.214&0.280 \\
&  $\hat{\gamma}_{2,1}$  & 0.781 &0.295& 0.358 &0.174&0.350&0.384 \\
\hline

  &  $\hat{\rho}_p$& 0.526 &0.754& 0.546 &0.561&0.664&0.787 \\
 &  $\hat{\tau}$& 0.398 &0.567& 0.403 &0.430&0.551&0.646 \\
Versicolor&  ${\hat\rho}_G$  & 0.546 &0.756& 0.551 &0.584&0.687&0.790 \\
&  $\hat{\gamma}_{1,2}$  & 0.533 &0.744& 0.542 &0.580&0.658&0.787 \\
&  $\hat{\gamma}_{2,1}$ & 0.523 &0.766& 0.559 &0.572&0.682&0.809 \\
\hline

 &  $\hat{\rho}_p$ & 0.457 &0.864& 0.281 &0.401&0.538&0.322 \\
 &  $\hat{\tau}$& 0.307&0.670& 0.219 &0.291&0.419&0.271 \\
Virginica&  $\hat{\rho}_G$  & 0.687 &0.820& 0.455 &0.621&0.623&0.519 \\
&  $\hat{\gamma}_{1,2}$   & 0.406 &0.867& 0.278 &0.467&0.567&0.304 \\
&  $\hat{\gamma}_{2,1}$ & 0.476 &0.832& 0.315 &0.308&0.548&0.355 \\
\hline\hline
\end{tabular}
\end{table}

From Table \ref{tab:5}, we observe that comparing with other two species, Iris Setosa has high correlation between sepal length and sepal width, but has low correlation between sepal length and petal length. Versicolor has much larger correlation between petal length and petal width than the other two species do. Virginica has the highest correlation between lengths of sepal and petal among the three species. 

Kendall's $\tau$ correlation value is the smallest among all correlation estimators across all pairs and across all species.  Two regular Gini correlation estimators are quite different especially between sepal width and petal length in Iris Virginica species. The difference is as high as 0.159. One might perform a hypothesis test on exchangeability of two variables by testing $\gamma_{1,2}=\gamma_{2,1}$ (Schechtman, Yitzhaki and Artsev, 2007). The p-value of the test is 0.0113, which serves as a strong evidence to reject exchangeability of two variables sepal width and petal length in Iris Virginica. 
We also observe that  $\hat{\rho}_G$ and $\hat{\rho}_p$ tend to have a same pattern across variable pairs and across species.  For example, for all six pairs of variables in Iris Setosa, $\hat{\rho}_G$ is large or small whenever $\hat{\rho}_p$ is large or small. In other words, the correlation ranking across variable pairs provided by the Pearson correlation is the same as the ranking by the proposed symmetric Gini correlation. However, such a pattern is not shared by  any two correlations from $\hat{\rho}_G$, $\hat \tau$, $\hat{\gamma}_{1,2}$ and $\hat{\gamma}_{2, 1}$. Also, values of $\hat{\rho}_G$ are larger than values of $\hat{\rho}_p$ in most cases.  

\section{Conclusion}
In this paper we propose symmetrized Gini correlation $\rho_g$ and study its properties.
The relationship between $\rho_g$ and $\rho$ is established when the scatter matrix, $\Sigma$,  is  homogeneous. The affine invariant version $\rho_G$ is also proposed to deal with the case when $\Sigma$  is  heterogeneous.  Asymptotic normality of the proposed estimators are established. The  influence function reveals that $\rho_g$ is more robust than the Pearson correlation while it is  less robust than the Kendall's $\tau$ correlation. Comparing with the Pearson correlation estimator, the regular Gini correlation estimator and the Kendall's $\tau$ estimator of $\rho$, the proposed estimators balance well between efficiency and robustness and provide an attractive option for measuring correlation. Numerical studies demonstrate that  the proposed estimators have satisfactory performance under a variety of situations. In particular, the symmetric Gini estimators are more efficient than the regular Gini estimators. This can be explained by the fact that the multivariate spatial rank used in the symmetrized Gini correlations takes more dependence information than the marginal ranks in the traditional ones. 

We comment that the symmetric Gini correlation $\rho_g$ is not limited to elliptical distributions.  Theorems 2, 4 and 5 also hold for any bivariate distribution with a finite first moment.  Under elliptical distributions, the linear correlation parameter $\rho$ is well defined and all the  four estimators  are Fisher consistent.  Hence their asymptotical variances are comparable and can be used for evaluating relative asymptotic efficiency among the estimators. 

The proposed symmetric Gini correlation has some disadvantages. Although its formulation is natural, the symmetric Gini loses an intuitive interpretation. It is more difficult to compute than the Pearson correlation, especially when $X$ and $Y$ are heterogeneous. In that case, an iterative scheme is required to obtain the affine invariant version of symmetric Gini correlation.  When applying the proposed measure, one may consider the trade-off among efficiency, robustness, computation and interpretability.

\end{document}